\documentclass{aa}
\include{epsf}

\begin{document}

\title{Dynamo action in turbulent flows}

\author{V. Archontis \inst{1}
\and S.B.F. Dorch\inst{2}
\and {\AA}. Nordlund\inst{2}  }

\offprints{V. Archontis --- vasilis@ll.iac.es}

\institute{
     Instituto de Astrofisica de Canarias, Via Lactea s/n E-38200,
     La Laguna, Spain
\and
     The Niels Bohr Institute for Astronomy, Physics and Geophysics,
     Juliane Maries Vej 30, DK-2100 Copenhagen {\O}, Denmark
}

\date{Received date, accepted date}

\authorrunning{Archontis et al.}

\def\Rm{{\rm Re}_{\rm m}}
\def\Re{{\rm Re}}
\def\Prm{{\rm Pr}_{\rm m}}
\def\Fkin{{\rm {\bf F}}_{\rm kin}}
\def\Fconv{{\rm {\bf F}}_{\rm conv}}
\def\Fpoynt{{\rm {\bf F}}_{\rm poynt}}
\def\fext{{\rm {\bf f}}_{\rm ext}}
\def\flor{{\rm {\bf f}}_{\rm L}}
\def\Ekin{{\rm E}_{\rm kin}}
\def\Emag{{\rm E}_{\rm mag}}
\def\Etot{{\rm E}_{\rm tot}}
\def\Etherm{{\rm E}_{\rm th}}
\def\Qvisc{{\rm Q}_{\rm v}}
\def\Qjoule{{\rm Q}_{\rm J}}
\def\Qcool{{\rm Q}_{\rm cool}}
\def\Wlor{{\rm W}_{\rm L}}
\def\uu{{\bf u}}
\def\bb{{\bf B}}
\def\jj{{\bf j}}
\def\ee{{\bf e}}
\def\BB{{\bf B}}
\def\BE{\begin{equation}}
\def\EE{\end{equation}}
\def\div{\nabla \cdot}

\abstract{We present results from numerical simulations of
nonlinear MHD dynamo action produced by three-dimensional flows
that become turbulent for high values of the fluid Reynolds number.
The magnitude of the forcing function driving the flow is allowed to evolve
with time in such way as to maintain an approximately constant
velocity amplitude (and average kinetic energy) when the flow
becomes hydrodynamically unstable. It is found that the saturation
level of the dynamo increases with the fluid Reynolds number (at
constant magnetic Prandtl number), and that the average growth
rate approaches an asymptotic value for high fluid Reynolds number.
The generation and destruction of magnetic field is examined
during the laminar and turbulent phase of the flow and it is found
that in the neighborhood of strong magnetic flux ``cigars"
Joule dissipation is balanced by the work done against the
Lorentz force, while the steady increase of magnetic energy
occurs mainly through work done in the weak part of the magnetic
field. \keywords{Physical data and processes:
magnetic fields --- MHD --- turbulence --- diffusion} }

\maketitle

\section{Introduction}

The dynamo problem is mainly related to the understanding of the
processes of magnetic field generation due to the motions of
conducting fluids, the so-called dynamo action (Cowling
\cite{Cowling1934}, Moffatt \cite{Moffatt1978}, Parker
\cite{Parker1979}). One often divides dynamo theory into two
theoretical regimes: First, the kinematic dynamo problem in which
magnetic fields and flows are decoupled. The velocity of the flow
is prescribed and the evolution of the magnetic field is governed
by the induction equation (here in dimensionless form):
\begin{equation}
\frac{\partial {\bf B}}{\partial t} = \nabla \times ( {\bf u} \times {\bf
B})
    + \frac{1}{{\rm Re}_{\rm m}} \nabla^2 {\bf B}, \label{induction.eq}
\end{equation}
where {\bf B} is the magnetic field and {\bf u} is the prescribed
velocity field. The  quantity ${\rm Re}_{\rm m}=u\ell/\eta$ is the
magnetic Reynolds number and it is often huge in astrophysical
systems, where $u$ and $\ell$ are characteristic velocity and
length scales and $\eta$ is the magnetic diffusivity. A flow acts
as a fast dynamo if the growth rate does not tend to zero in the
limit of infinite magnetic Reynolds number. The solar dynamo is an
astrophysical example of fast dynamo action since it operates on
the convective time scale of the fluid which is fast compared to
that of the Ohmic diffusion. In the fast kinematic dynamo problem
one seeks velocity fields that lead to the exponential
amplification of the magnetic field through induction Eq.\
(\ref{induction.eq}). A powerful mechanism for growth of the
magnetic field is the stretching of the magnetic field lines
(cf.\ Archontis et al.\ \cite{Archontis+ea03}), and
is primarily achieved by chaotic flows. The kinematic approach is
valid when the magnetic field is weak and there are no further
dynamical effects.
{ In the second, fully dynamical dynamo regime, the velocity
field is not prescribed and the exponential growth of the magnetic
field saturates when the field becomes strong enough to modify the
flow sufficiently through the feedback by the Lorentz force.
The saturation state of the dynamo then emerges as a self consistent
solution to the equations of nonlinear magneto-hydrodynamics.
In a fully self-consistent dynamical experiment the forcing would
have internal causes, e.g.\ of convective origin---here we are
concerned with the case of an externally prescribed forcing function.}

The nonlinear properties of fast dynamos have received considerable
attention and a variety of both analytical and numerical studies
have provided a valuable insight into the nature of the dynamo
action and saturation (Nordlund et al.\ \cite{Nordlund+ea92};
Brandenburg et al.\ \cite{Brandenburg+ea95}; Cattaneo et al.\
\cite{Cattaneo+ea96}; Zienicke et al.\ \cite{Zienicke+ea98};
Brummell et al.\ \cite{Brummell+ea98}). However, a difficulty
that arises in numerical simulations is that often, fast dynamo
flows become hydrodynamically unstable at fluid Reynolds number
(Re$=u\ell/\nu$, $\nu$ being the viscosity) greater than a
critical value (Podvigina \& Pouquet \cite{Podvigina+Pouquet1994})
and spatio-temporal turbulence appears to be nascent for higher
values of Re. As a result the desired flow velocity is modified
even in the kinematic regime and the problem contains the
complexity of that of a turbulent dynamo. Thus, the choice of the
forcing function that drives the flow is crucial. We use a forcing
function with an amplitude that is allowed to evolve with time,
keeping the average kinetic energy approximately
constant through both laminar and turbulent phases.
The growth rates and the saturation level of such
turbulent dynamos are examined with respect to the magnetic and
fluid Reynolds numbers. Also of great interest is the
understanding of the processes at work in the kinematic (both
laminar and turbulent phase of the flow) and in the saturated
regime. We present an analysis of the Lorentz work and Joule
dissipation, which shows that they are in close balance and that
dynamo-work occurs mainly in the weak part of the magnetic field
where dissipation is insignificant.

The paper is organized as follows. Section \ref{simulation.sec}
contains the equations, the numerical method and a description
of the driving of the flow that has been used in the simulation.
Results for turbulent dynamo action are presented in Section
\ref{results.sec}. The structures of the velocity field and the
magnetic field in physical space are discussed in Section
\ref{patterns.sec}. Whether or not the growth rates and the
saturation level of the dynamo depend on the magnetic or fluid
Reynolds number is examined in Section \ref{saturation.sec} and
Section \ref{growth.sec}. A qualitative understanding of the
nature of the dynamo is discussed in Section \ref{nature.sec},
using results from a kinematic dynamo experiment.
Section \ref{conclusions.sec} contains the overall conclusions of
the present numerical simulations.

\section {The simulation}
\label{simulation.sec}

\subsection{The equations}

The compressible MHD equations are solved numerically in a
periodic computational domain, with periodicity of ${2\pi}$ in all
three directions. Apart from the magnetic induction equation Eq.\
(\ref{induction.eq}), the additional equations solved are the
following:
\begin{eqnarray}
{\frac{\partial {\bf \rho}} {\partial t}} & = & -\nabla\cdot\rho{\bf u},\\
\label{mass.eq}
{\frac{\partial {({\rho}{\bf u})}} {\partial t}}& = & -\nabla P
    + {\bf j}\times{\bf B} + {\bf f} - \nabla\cdot(\rho{\bf u}{\bf u}) \label{motion.eq},\\
{\frac{\partial e}  {\partial t}} & = & - \nabla\cdot(e{\bf u})
    - P\nabla\cdot{\bf u} + \Qvisc + \Qjoule + \Qcool \label{energy.eq}
\end{eqnarray}
where $\rho$ is the fluid density, $P$ the pressure, ${\bf j}$ is
the electric current density, $e$ is the internal energy
and ${\bf f}$ is an external forcing term.
$\Qvisc$ and $\Qjoule$ are the viscous and Joule dissipation
respectively. In the experiments, a Newtonian cooling term,
$\Qcool = (T-T_0)/\tau_{\rm cool}$ is used, where $T$ is the
temperature and $\tau_{\rm cool}=5$. { The cooling term only works
as a sink of the thermal energy, in order to balance the heat
generation by dissipative terms in the energy equation and the
actual magnitude of the cooling time is of little importance
with no effect on the results of our simulation.}

The above equations are solved numerically on a staggered mesh.
The time stepping is performed by a third
order predictor-corrector method (Hyman \cite{Hyman1979}). The
derivatives and interpolations are of 6th and 5th order
respectively and the numerical scheme conserves $\nabla\cdot{\bf
B}=0$ exactly. Numerical solutions are obtained on a grid of a
maximum of $180^{3}$ points, using a modified version of the code
by Galsgaard, Nordlund and others (Galsgaard \& Nordlund
\cite{Galsgaard+Nordlund1997}; Nordlund et al.
\cite{Nordlund+ea92}). The initial magnetic field is chosen to
be a weak random perturbation with an amplitude of $ 10^{-5}$
in non-dimensional units.

\subsection{Flow considerations}

There is a considerable body of work in the literature that deals
with fast dynamo action in simple, steady and three-dimensional
flows. In our simulation the velocity is chosen to have vigorous
(fast) dynamo properties in the kinematic regime, where the
Lorentz force is negligible. One such class of flows, which are
well known candidates for fast dynamo action is the ABC flows. The
{form of the velocity} is given by
\begin{eqnarray}
{\bf u}_{\rm ABC} & = & A(0,\sin kx, \cos kx) + B(\cos ky,0,\sin ky)\\ \nonumber
    & +  & C(\sin kz,\cos kz,0)
\label{flowform.eq}
\end{eqnarray}
This periodic flow is the sum of three steady Beltrami waves, parameterized
by $A$, $B$ and $C$ and has the property $\nabla\times {\bf u}
\propto {\bf u}$ (velocity is parallel to vorticity). If
one or more of the constant coefficients ($A$, $B$ and $C$) is
zero, the flow is integrable and is not a fast dynamo. If all
three are non-zero the flow is non-integrable (Dombre et al.
\cite{Dombre+ea86}) and contains a mixture of chaotic regions and
regular islands. The special case $A=B=C=1$ was introduced by
Childress (\cite{Childress+ea70}) as a model for kinematic dynamo
action. Arnold et al. (\cite{Arnold+ea83}) noted that a steady
three-dimensional flow with infinite conductivity and chaotic
streamlines favors the growth of magnetic fields. Their results
were followed by the work of Galloway et al.
(\cite{Galloway+ea84}) and Moffatt et al. (\cite{Moffatt+ea85})
who studied dynamo action in ABC flows with finite conductivity. A
detailed description of the amplification process responsible for
the kinematic dynamo action, when the wavenumber of the flow $k$
is equal to one or higher, has been provided by
Dorch (\cite{Dorch2000}) and Archontis et al. (\cite{Archontis+ea03}).
The hydrodynamic stability of the ABC flow has been studied by
Podvigina \& Pouquet (\cite{Podvigina+Pouquet1994}) who found that
for values of fluid Reynolds number above a critical one (${\rm
Re}_{\rm c}=13$) the ABC flow destabilizes, first to time
dependent but still smooth states, and then to a turbulent state.
{In our simulation, the definition of the Reynolds numbers is the same
as in Galanti et al. (\cite{Galanti+ea92}).}

In the following experiments the initial flow is taken to be
an ABC flow with $A=0.9$, $B=1$, $C=1.1$ and wavenumber equal to
unity. The reason for choosing values of $A$,
$B$, and $C$ slightly different from unity,
is that the $A=B=C$ case {is too special, because
of its exact symmetry}, and thus turbulence takes longer to develop. The
purpose of our work is to study dynamo action in the turbulent
regime of driven ABC flows since astrophysical dynamos typically
occur in turbulent environments with very high fluid and magnetic
Reynolds numbers. However, the behavior of turbulent dynamos in
the high Re, ${\rm Re}_{\rm m}$ limit may be expected to be to
some extent generic. {An increase in the Reynolds numbers involves
smaller and smaller length scales where dissipation actually occurs.
However, the rate of dissipation is determined by the large scale
developments of the flow and is independent of the magnitude of
the viscosity once the Reynolds numbers are large.  Thus, there is
good hope that also dynamo properties are asymptotically independent of
the Reynolds numbers.} That hope is indeed borne out by the present
numerical experiments, as is discussed in more details below.

\section{Results}
\label{results.sec}

As a starting point, the forcing function which is imposed is such
that the ABC flow Eq.\ (\ref{flowform.eq}) appears. The forcing
necessary to accomplish this {must have} the following form:
\begin{equation}
{\bf f} = -\frac{1}{{\rm R}_{\rm e}} \nabla^{2}{\bf u}_{\rm ABC}
\label{forcing.eq}
\end{equation}
However, in order to ensure that the kinetic energy is maintained
at a value close to the initial value, the amplitude of the
forcing is controlled by a differential equation, which
compensates for the increased dissipation in the turbulent phase
by increasing the forcing when the kinetic energy tends to drop
below the nominal value. The amplitude factor in front of the
driving force is given by
\begin{equation}
K = {{\rm E}_0 \over {\rm E}_{\rm kin}} e^L, \label{factor.eq}
\end{equation}
where $L$ is determined from
\begin{equation}
{dL \over dt} = {1\over\tau_{\rm t.o.}} ln {{\rm E}_0\over {\rm E}_{\rm
kin}} ,
\end{equation}
with $L(t=0) = 0$. The differential equation evolves the amplitude
factor $e^L$ on the turn over time scale, $\tau_{\rm t.o.}$, if
the kinetic energy tends to deviate from the nominal value ${\rm
E}_0$. The immediate factor ${\rm E}_0/{\rm E}_{\rm kin}$ helps to
reduce the time delay that results from the integral nature of the
exponential factor, and the delayed response of the kinetic energy
on changes in the driving. The kinetic energy control is an
essential feature, in that otherwise the amplitude of the velocity
in the turbulent phase would become arbitrarily small, at
increasingly large fluid Reynolds number, since the forcing in the
laminar phase needs to scale with $1/{\rm Re}$.  It is indeed
necessary to maintain the kinetic energy in order to maintain the
same turn over time and actual fluid Reynolds number, since the
formal fluid Reynolds number is defined in terms of the RMS
velocity of the laminar ABC flow.

\begin{figure}[!htb]
\makebox[8cm]{
\epsfxsize=6.0cm
\epsfysize=5.0cm
\epsfbox{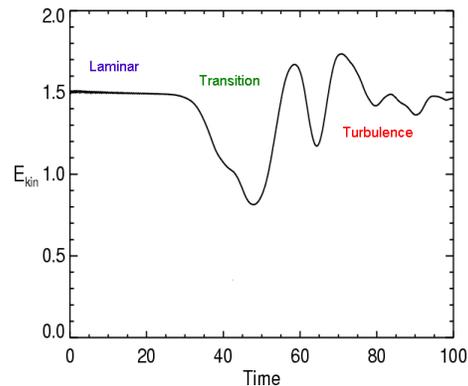}
}
\caption[]{\small Temporal evolution of the kinetic energy showing the
transition
from a laminar to a turbulent state.}
\label{fig1}
\end{figure}

Figure \ref{fig1} shows the evolution of the kinetic energy in
time for the first 100
time units. Two phases may be identified. The laminar phase which extends
up to $t=30$ and the turbulent regime for $t>30$. Note that the kinetic
energy drops by up to about 40\% when the flow enters the
turbulent phase. After a short adjustment time, the forcing control
returns the kinetic energy to the nominal level, where it is then
maintained, except for minor fluctuations.

\begin{figure}[!htb]
\makebox[8cm]{
\epsfxsize=7.0cm
\epsfysize=6.0cm
\epsfbox{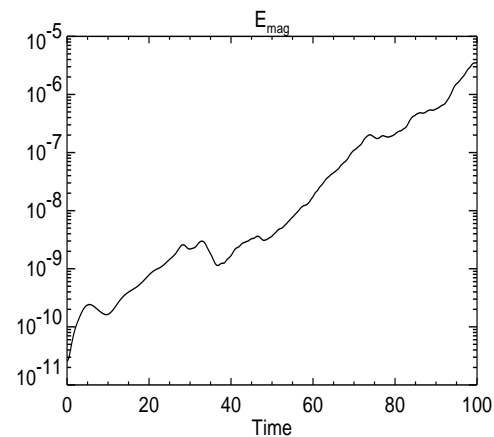}
}
\caption[]{\small Temporal evolution of the magnetic energy during the
kinematic
regime of the dynamo.}
\label{fig2}
\end{figure}

The magnetic energy for the
same time period during the laminar phase increases
exponentially and the growth rate is about 0.15.
When the flow enters the turbulent phase ($t>30$)
the magnetic energy drops slightly and later ($t=40$)
continues to grow exponentially with a similar growth rate as in the
laminar phase (Fig.\ \ref{fig2}).

\begin{figure}[!htb]
\makebox[8cm]{
\epsfxsize=7.0cm
\epsfysize=6.0cm
\epsfbox{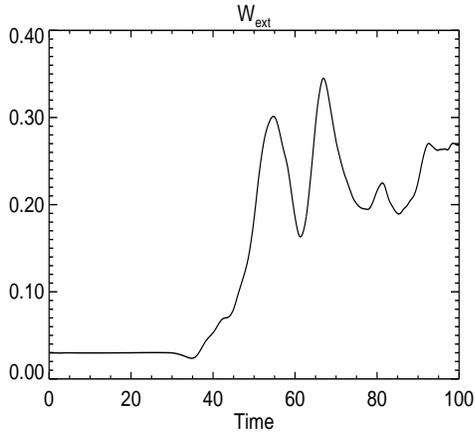} }
\caption[]{\small The evolution of the external average work in time.}
\label{fig3}
\end{figure}

An analysis of the external work in the above two distinct phases
shows that the average work in the turbulent regime is much larger
than in the laminar case (Fig.\ \ref{fig3}). On the first
hand, the asymptotic level of the average work ($\approx 0.25$) is
independent of Re and incidentally corresponds closely to the
level that would be obtained for the critical value (${\rm
Re}_{\rm c}$). On the other hand, the average work in the laminar
regime is proportional to $1/{\rm Re}$, and can become arbitrarily
small for high enough Reynolds numbers.

This illustrates the point made above, about the forcing control;
without it, the
amplitude of the velocity in the turbulent phase would drop very much
and the kinetic energy could not be maintained in order to have
constant fluid Reynolds number during the simulation.

\begin{figure}[!htb]
\makebox[8cm]{
\epsfxsize=7.0cm
\epsfysize=6.0cm
\epsfbox{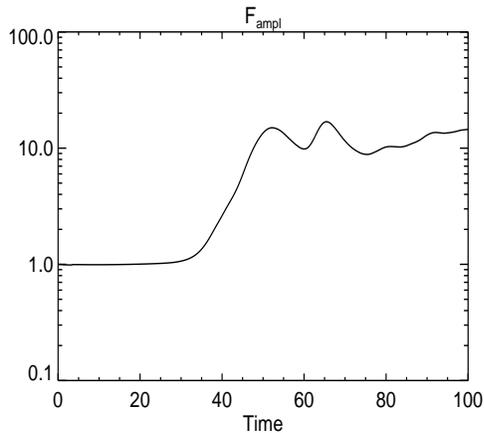} }
\caption[]{\small The increase of the amplitude of the driving force going
from the laminar phase to the turbulent phase.}
\label{fig4}
\end{figure}

The enhancement of the driving force above the laminar value
is shown in Fig.\ \ref{fig4}. The factor in Eq.\ (\ref{factor.eq})
is by definition $=1$ during the
laminar phase of the flow but increases during the turbulent phase to
maintain the kinetic energy.
The asymptotic level of the factor is about 10 for ${\rm Re}=100$.

\begin{figure}[!htb]
\makebox[8cm]{
\epsfxsize=7.0cm
\epsfysize=6.0cm
\epsfbox{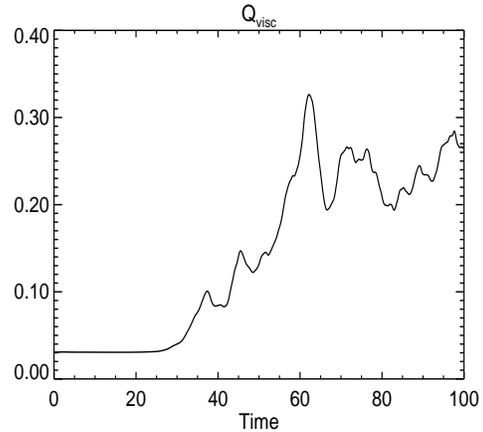} }
\caption[]{\small  The evolution of the viscous dissipation up to $t=100$.}
\label{fig5}
\end{figure}

Figure \ref{fig5} shows the viscous dissipation as a function of time
for the laminar and turbulent phase. During the laminar phase the
dissipation is constant and is balanced by the external work.
The viscous dissipation increases when
the flow becomes turbulent. It reaches an asymptotic level when the
turbulent velocity is maintained at the same level as that of the
laminar flow.

\begin{figure*}[!htb]
\begin{center}
\makebox[16cm]{
\epsfxsize=7.0cm
\epsfysize=6.0cm
\epsfbox{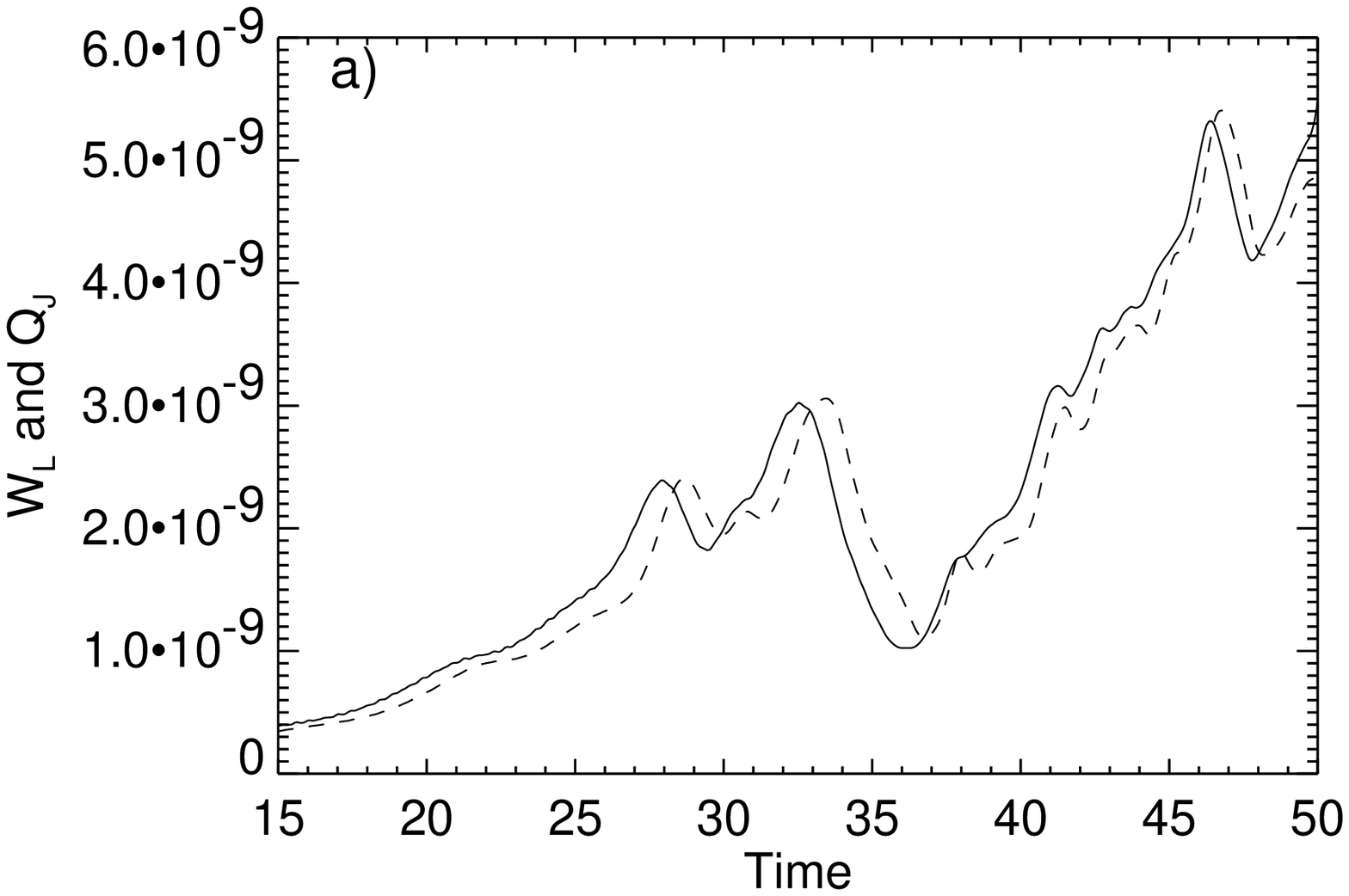}
\epsfxsize=7.0cm
\epsfysize=6.0cm
\epsfbox{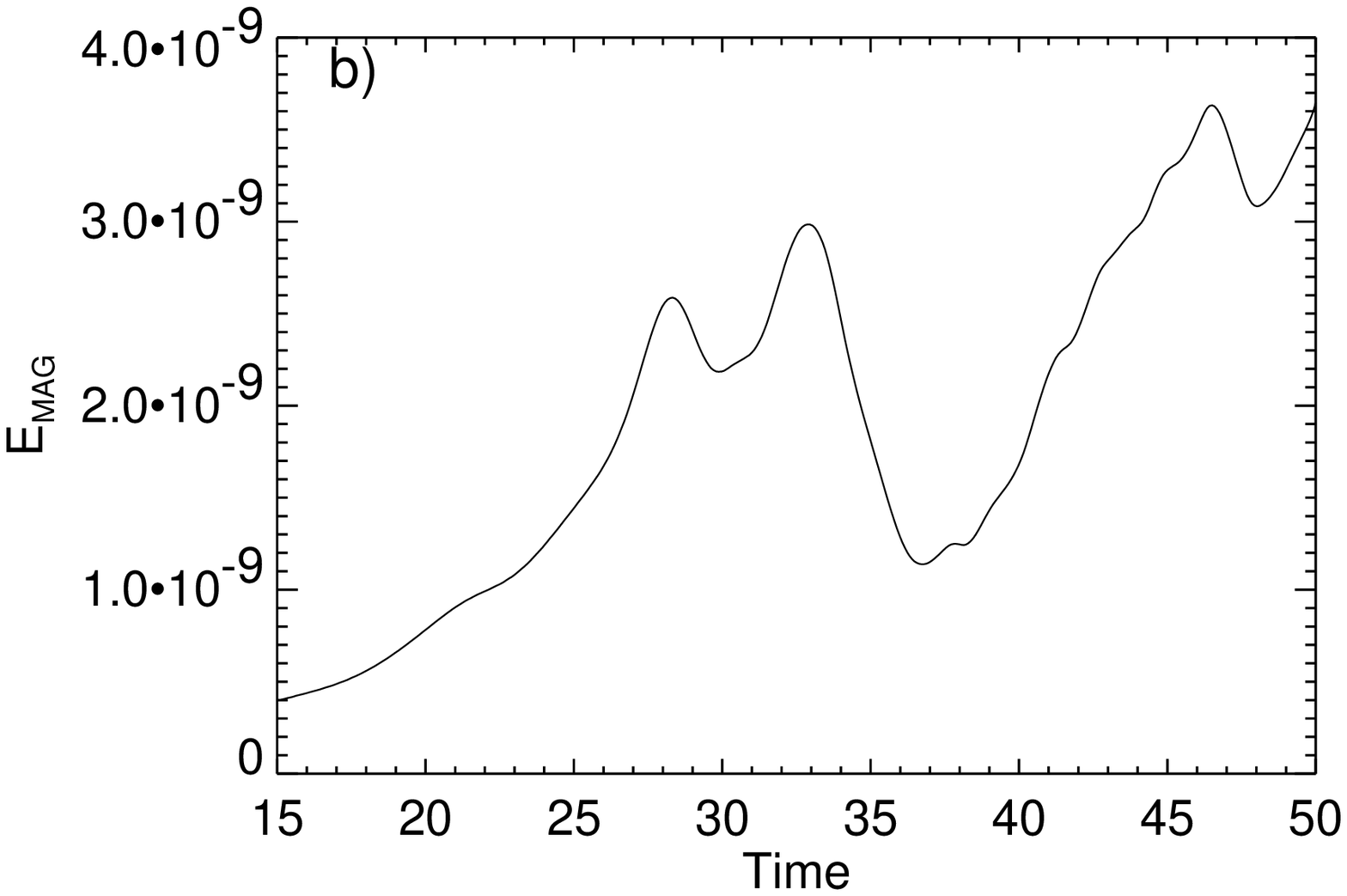} }
\makebox[16cm]{
\epsfxsize=7.0cm
\epsfysize=6.0cm
\epsfbox{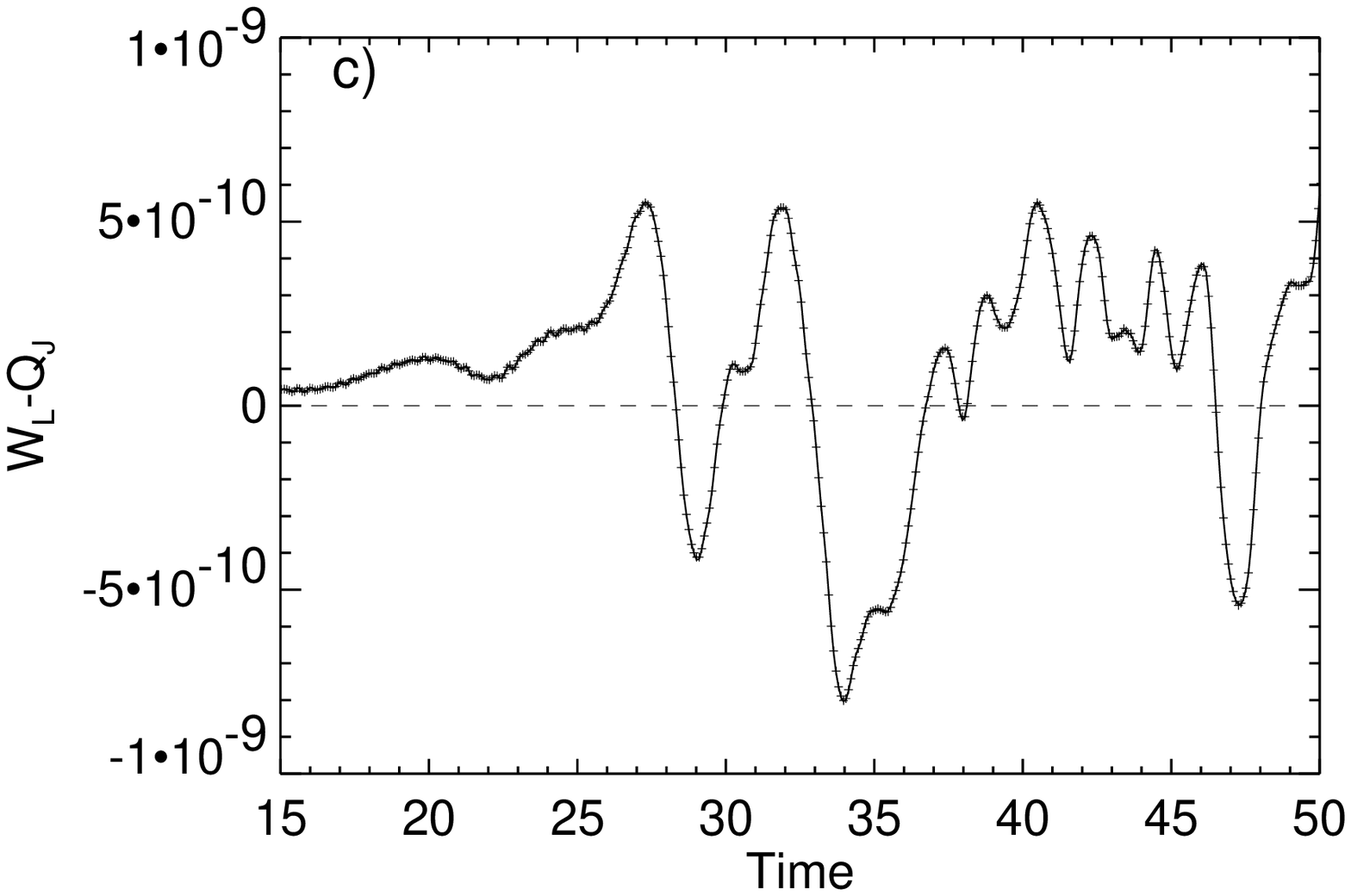}
\epsfxsize=7.0cm
\epsfysize=6.0cm
\epsfbox{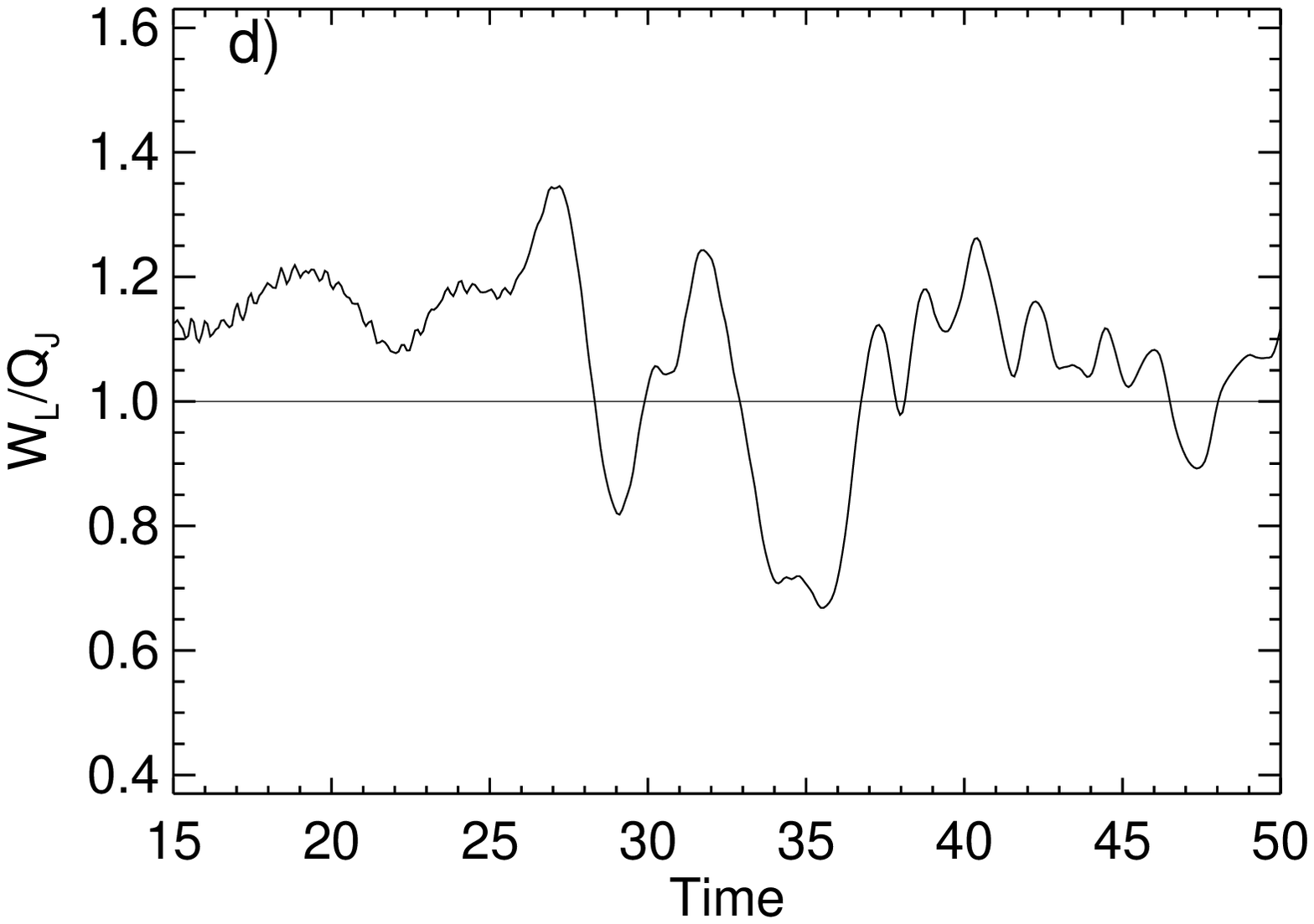} }
\end{center}
\caption[]{\small  Four panels showing different aspects of the
time evolution of the Lorentz
work and Joule dissipation during the linear regime of the dynamo and
comparison with the magnetic energy:
{\bf a}) The Lorentz work W$_{\rm L}$ (full curve) and the Joule
dissipation Q$_{\rm J}$ (dashed curve).
{\bf b}) Total magnetic energy E$_{\rm mag}$.
{\bf c}) The difference between the Lorentz work and the Joule dissipation.
{\bf d}) Ratio between work and heating.
}
\label{fig6}
\end{figure*}

Figure \ref{fig6} compares the Joule dissipation to the work done
against the Lorentz force and also shows the temporal evolution of the magnetic
energy for the kinematic regime of the dynamo.

The work done against the Lorentz force is leading the Joule
dissipation and is ultimately responsible for the increase in the
magnetic energy. There is a close balance
between them with a small positive difference during the increase of the
magnetic energy and small negative difference when the magnetic
energy decreases, see Fig.\ \ref{fig6}a, Fig.\ \ref{fig6}b.
The fluctuations of the Lorentz work lead similar fluctuations of the
Joule dissipation when the magnetic energy increases or decreases.
Their difference in the saturated state of the dynamo fluctuates
around zero, as expected. The work done by the fluid on the
magnetic field is, on average, converted into Joule heat and
none is left to increase the magnetic energy.

The accuracy in the numerical method is demonstrated in Fig.\
\ref{fig6}c, where the $d{\rm E}_{\rm mag}/dt$ is
overplotted (symbols) on top of the difference ${\rm W}_{\rm
L}-{\rm Q}_{\rm J}$. These are averages over the periodic box, for
which the flux term in the magnetic energy equation vanish.
The magnetic energy equation is
\BE
{\partial {\rm E}_{\rm mag} \over \partial t} =
- \div {\rm {\bf F}}_{\rm poynt} - \uu\cdot\flor - \Qjoule. \label{emag.eq}
\EE
The source of magnetic energy is the work done
against the Lorentz force and the sink is the Joule dissipation.
Magnetic energy is transported by the Poynting flux, ${\rm {\bf
F}}_{\rm poynt} = {\bf E} \times {\bf B}$. According
to Eq. (\ref{emag.eq}), $d{\rm E}_{\rm mag}/dt$ should be identically equal
to ${\rm W}_{\rm L}-{\rm Q}_{\rm J}$, and the two are
indeed very nearly the same.

Note that the ratio of the Lorentz work to the Joule dissipation
in Fig.\ \ref{fig6}d
remains close to unity both during intervals of
time when the magnetic energy is growing and during intervals of
time when the magnetic energy is decaying.

\section{Magnetic and velocity patterns}
\label{patterns.sec}

During the laminar phase the flow has points with two-dimensional
stable manifolds (so called $\alpha$-type stagnation points),
where the magnetic field is concentrated along magnetic ``flux
cigars''. There are also points with two-dimensional unstable
manifolds, similar to the $\beta$-type stagnation points of the
{A=B=C=1 flow}, where magnetic flux sheets are formed (see
Dorch \cite{Dorch2000}).

\begin{figure*}[!htb]
\begin{center}
\makebox[16cm]{
\epsfxsize=8.0cm
\epsfysize=8.0cm
\epsfbox{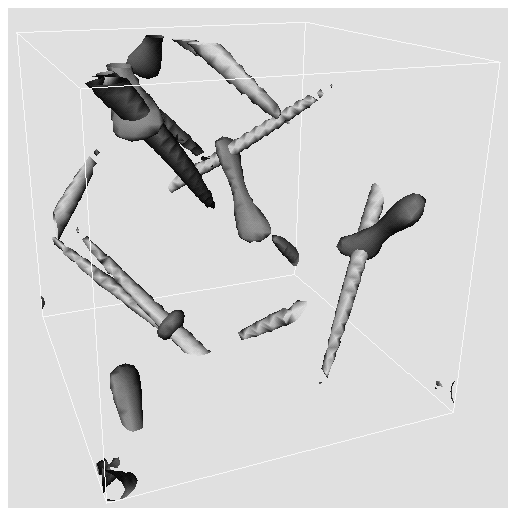}
\epsfxsize=8.0cm
\epsfysize=8.0cm
\epsfbox{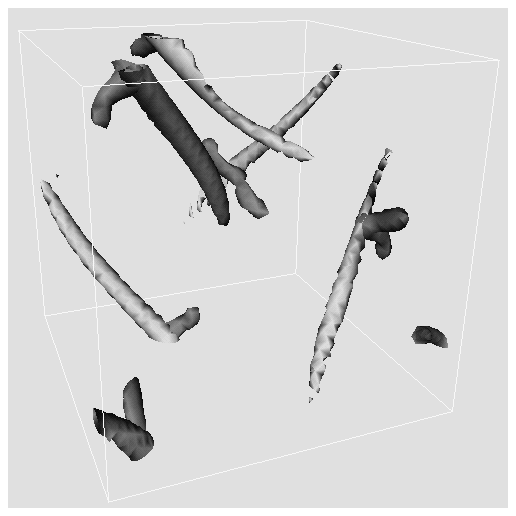} }
\caption[]{\small Isosurfaces of magnetic field strength (light) and {low}
velocity (dark) at $t=20$ (left panel) and $t=30$ (right panel). The
isosurface level of the magnetic field strength is $75\%$ of the peak
value in the snapshot.  {The velocity isosurfaces correspond to $4\%$
of the peak velocity in each snapshot.}}
\label{fig7}
\end{center}
\end{figure*}

Figure \ref{fig7} (left panel) shows that flux
cigars form an inclined triangle around a $\beta$-type
stagnation point (located at the center of the box). Another
one, which is located at the upper left corner, is pointing
towards the plane of the $\beta$-type point {while the flow
is still laminar}. The right panel
of the same figure shows {low velocity and high magnetic
field} isosurfaces at $t=30$. The velocity field comes into a
turbulent phase and the regions of {low velocity} are disturbed,
but the triangle formed by the flux cigars is still visible.
{The low velocity isosurfaces no longer contain points of exact
stagnation, but still correspond to regions of vigorous stretching
(see Archontis et al. \cite{Archontis+ea03}), which is why we
choose to visualize them.}

\begin{figure}[htb]
\begin{center}
\makebox[8cm]{
\epsfxsize=8.0cm
\epsfysize=8.0cm
\epsfbox{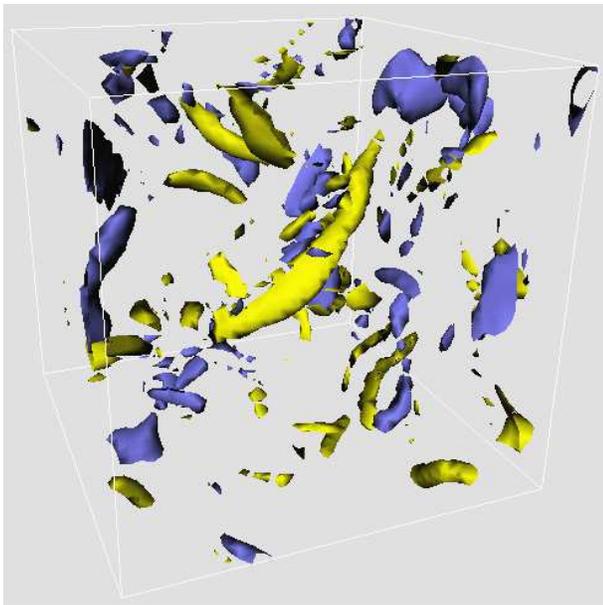} }
\caption[]{\small Isosurfaces of magnetic field strength (light) and {low
velocity} (dark) at $t=80$.}
\label{fig8}
\end{center}
\end{figure}

Figure \ref{fig8} is a visualization of the same velocity and
magnetic field strength for $t=80$. The velocity field has
a more turbulent configuration and the regions with velocities
less that $4\%$ of the peak
have disappeared. The magnetic flux cigars also disappear
and the magnetic energy is confined into curved tubes and weaker sheets.

\section{Saturation of the dynamo}
\label{saturation.sec}

Two experiments were performed to show the dependence of the
saturation level of the turbulent dynamo on the magnetic and fluid
Reynolds number. Cases with ${\rm Re}_{\rm m}={\rm Re}=200$ and
${\rm Re}_{\rm m}={\rm Re}=400$ are studied and the temporal
evolution of the magnetic energy is followed in the kinematic and
saturated regime.

\begin{figure}[!htb]
\makebox[8cm]{ \epsfxsize=8.0cm \epsfysize=6.0cm \epsfbox{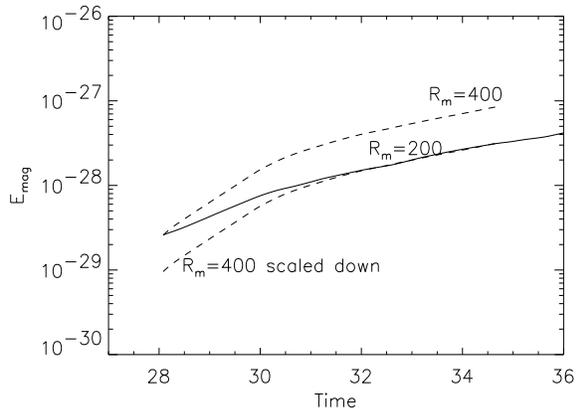}
} \caption[]{\small The growth of the magnetic energy in the
kinematic regime when ${\rm Re}_{\rm m}={\rm Re}=200$ (straight
line) at $80^{3}$ resolution and ${\rm Re}_{\rm m}={\rm Re}=400$
(dashed line) at $160^{3}$ resolution. The latter divided by 2.7
is shown by the lower (dashed) line.} \label{fig9}
\end{figure}

Figure \ref{fig9} shows the growth of the magnetic energy
in the beginning of the kinematic regime for ${\rm Re}_{\rm
m}={\rm Re}=200$ and ${\rm Re}_{\rm m}={\rm Re}=400$. The
numerical resolution used were $80^{3}$ and $160^{3}$,
respectively. After an initial transient phase the growth rate for
the second experiment is very similar to the growth rate of the
first one. The difference in the magnetic energy amounts to a
factor of 2.7. Thus if the magnetic energy for the $160^{3}$
experiment is divided with 2.7, it falls right on top of the
$80^{3}$ experiment with ${\rm Re}_{\rm m}=100$ (see Fig.\
\ref{fig9}).

Keeping the same values of magnetic and fluid Reynolds number, the
temporal evolution of the magnetic energy is examined in the
saturated regime. The magnetic energy eventually stops growing
exponentially and saturates. A result of considerable significance
is that for higher values of the fluid Reynolds number,
but constant magnetic Prandtl number $({\rm Pr}_{\rm
m}=\nu/\eta)$, the saturation level of the turbulent dynamo
increases, rather than decreases (Vainshtein \& Cattaneo \cite{Vainshtein+Cattaneo1992}).

\begin{figure}[!htb]
\begin{center}
\makebox[8cm]{ \epsfxsize=8.0cm \epsfysize=6.0cm
\epsfbox{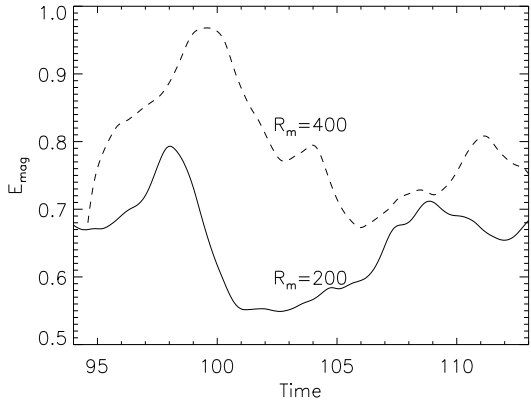} } \caption[]{\small Temporal evolution  of the
magnetic energy in the saturated regime for ${\rm Re}_{\rm
m}=200$, ${\rm Re}=200$ (solid line) and ${\rm Re}_{\rm m}=400$,
${\rm Re}=400$ (dashed line). } \label{fig10}
\end{center}
\end{figure}

Figure \ref{fig10} shows the temporal evolution of the magnetic
energy over a small epoch of the non-linear regime, illustrating the
effect of increased numerical resolution. There is an
initial transient, as the larger resolution allows field
concentrations to collapse further. Note that
there are then sections of time where the growth rates and the
decay rates are quite similar in the ${\rm Re}_{\rm m}=400$
and ${\rm Re}_{\rm m}=200$ cases.
The sections after the largest maximum in the two cases, and the rise
towards the next maximum are two examples. This might be fortuitous,
but more likely it corresponds to parallel evolution of similar
magnetic structures.

The overall level is slightly higher, and the evolution is
slightly delayed in the high ${\rm Re}$ case, because flux
structures that were barely resolved at $80^3$ are allowed to
collapse in the $160^3$ case. This is indeed also what happens in
the kinematic case, but here the size of most structures is
controlled by the Lorentz force, and not by diffusion, so the
effect is only marginal. For the same reason, the effect may be
expected to vanish altogether as the magnetic Reynolds number goes
to infinity. {In addition, the size of the magnetic structures in the kinematic
regime scales as ${\rm Re}_{\rm m}^{-1/2}$. In the saturation
regime the size is again determined by the balance between the
advection and dissipation of the magnetic energy and presumably
changes following the same scaling with no crucial dependence on the
value of ${\rm Re}$.}

After some time, the saturation level for ${\rm Re}_{\rm m}=400$, indeed
fluctuates around a level that is only marginally larger than
for ${\rm Re}_{\rm m}=200$. This is small compared to the factor
2.7 for the kinematic regime of the dynamo.

One also notices that there are more fluctuations in the higher
resolution case (e.g.\ the extra bump on the way down after the
largest maximum), as a result of the small scales resolved
when going to higher ${\rm Re}_{\rm m}$.
That trend is expected to continue with increasing magnetic
Reynolds number.

\section{Magnetic Prandtl number and growth rate}
\label{growth.sec}

In Section \ref{saturation.sec} we examined the growth rate
and the saturation level of a fast dynamo when the
magnetic Prandtl number is equal to one.
Equally important is the study of the degree of their
dependence when ${\rm Pr}_{\rm m}$ is varied. Thus,
additional experiments were performed where the forcing again
is such that it maintains the kinetic energy at a value close to the
initial one, apart from small oscillations.

\begin{figure}[!htb]
\begin{center}
\makebox[8cm]{ \epsfxsize=8.0cm \epsfysize=6.0cm
\epsfbox{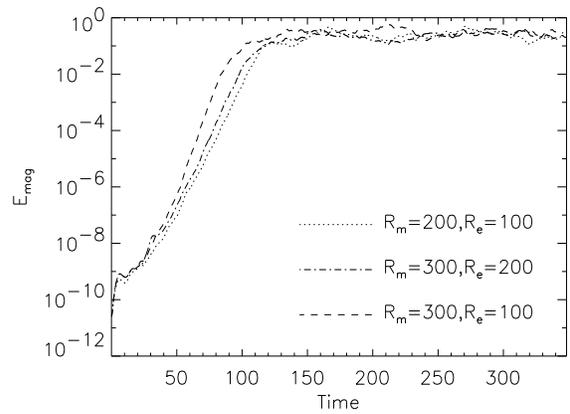} } \caption[]{\small Temporal evolution of the
magnetic energy for ${\rm Pr}_{\rm m} \geq 1$. Numerical resolution
is $96^3$ for ${\rm Re}_{\rm m}=300$ and $80^3$ for
${\rm Re}_{\rm m}=200$.}
\label{fig11}
\end{center}
\end{figure}

\begin{figure}[!htb]
\begin{center}
\makebox[8cm]{ \epsfxsize=8.0cm \epsfysize=6.0cm
\epsfbox{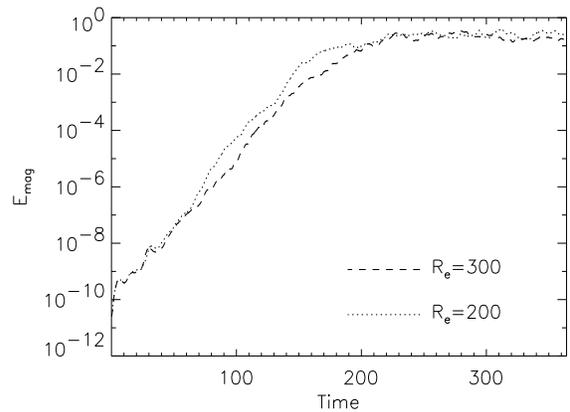} } \caption[]{\small Temporal evolution of
the magnetic energy for ${\rm Re}_{\rm m}=200$ and
${\rm Pr}_{\rm m} \leq 1$.} \label{fig12}
\end{center}
\end{figure}

Not surprisingly, the growth rates during the laminar phase ($t<30$) of the
flow are found to be the same (Figures \ref {fig11}, \ref {fig12}).
By construction, the velocity is exactly the same in the laminar
phase, independent of Re, and the effect of increasing ${\rm Re}_{\rm m}$
is simply, as in the comparison of the ${\rm Re}_{\rm m}=200$ and
${\rm Re}_{\rm m}=400$ experiments, to give a different transient at
the beginning of the simulation.

Less trivially, the average growth rates in the turbulent regime also
turn out to be {similar} with no crucial dependence on the magnetic
Prandtl number for high ${\rm Re}$. This is illustrated
in Figure \ref {fig12}, where the difference in the magnetic
energy during the turbulent phase ($t>30$) is very small between
the two experiments with ${\rm Re}=200$ and ${\rm Re}=300$.
This may also be seen as a consequence of the generic
properties of high Reynolds number turbulence.
More specifically, by one of Kolmogorov's main assumptions,
energy dissipation (and hence the forcing) is independent
of the fluid viscosity once the Reynolds number is high enough.
Thus, and by analogy with the previous experiments, we expect
that the only important change in the magnetic field is the
development of more fine structure with increasing magnetic
Reynolds number. The growth rate is determined by the large scale
properties of the flow, and in particular by the turn over time,
which by construction does not change between these experiments.

Finally, the above experiments reinforce the conclusion from
Section \ref{saturation.sec}, that the saturation level of the dynamo
does not depend crucially on the value of ${\rm Pr}_{\rm m}$,
but only increases slightly with ${\rm Re}_{\rm m}$ as it is shown in
Figure \ref {fig11}.

\section{Nature of the dynamo}
\label{nature.sec}

{The results of the above non-linear experiments are similar to the results
of previous studies of turbulent hydromagnetic convection
(Nordlund et al.\ \cite{Nordlund+ea92})
and both results indicate that during the dynamo action} there is
a close balance between the work done against the Lorentz force
and Joule dissipation with a relatively small
positive (negative) difference when the magnetic energy increases
(decreases).  In other words, work and dissipation are in near
balance, with a relatively small difference responsible for the
average growth of the magnetic energy.  One may rightly ask why
this is so; is the small positive difference then just fortuitous,
and might the difference not just as well be negative for a slightly
different flow?

To address this question we make use of a kinematic
dynamo experiment, where the analysis is simpler than in a turbulent
case. Turbulence makes
the analysis of the Lorentz work  difficult---there is much time
dependent noise, which corresponds to waves averaging out over
time. The work is much less noisy when the velocity is prescribed
{and/or} has not yet entered the turbulent phase, and
we therefore analyze the results of an experiment that deals with
kinematic dynamo action. The velocity field is chosen to be the
{$A=B=C=1$} flow with $k=1$ and the magnetic Reynolds number
is ${\rm Re}_{\rm m}=300$.

The convergence and the diffusion rates at the points of strong
flux concentration are of the order of $\sqrt{2}$, which is much
larger than the dynamo growth rate (which is about 20 times
smaller, e.g.\ see Childress \& Gilbert \cite{STF}), so the
advection and diffusion are in close balance in the neighborhood
of the ``flux cigars". During the exponential growth of the
magnetic energy the size of the flux cigars does not change
dramatically, but the magnetic energy increases gradually.

A detailed analysis for the latter case confirms
that this is indeed the case, and that the bulk part of the Joule
dissipation and the work done against the Lorentz force are in
detailed balance. This is understandable as a balance between
kinetic energy being converted into magnetic energy, that is then
immediately converted into heat through Joule dissipation.

Significant net
positive work (defined by the difference ${\rm W}_{\rm L} - {\rm Q}_{\rm J}$)
occurs instead in regions where the magnetic field is weak,
and the dissipation is very weak as well. In these regions, which
cover 90\% of the volume in this example, there is almost ``pure
work"; the Joule dissipation is much smaller than the work, and
the net work is almost identical to the net work in the whole box.

In other words:  The dissipation is strong only in the neighborhood
of the $\alpha$-type stagnation
points, but there it is balanced almost exactly by
{advection/convergence.}

\begin{table}[htb]
\begin{center}
\begin{tabular}{|c|c|} \hline\hline
Average Lorentz work                    &       $326.3$   \\ \hline
Average Joule dissipation               &       $271.1$   \\ \hline
Average magnetic energy                 &       $332.3$   \\ \hline
Energy growth rate                      &       $0.166$   \\ \hline
Fractional dissipation level            &       $0.0012$  \\ \hline
Fractional volume                       &       $0.89$    \\ \hline
Net work in weak dissipation regime     &       $55.1$    \\ \hline
Net work in high dissipation regime     &       $0.09$    \\ \hline
\end{tabular}
\caption[]{\small Summary of average values from the kinematic experiment with
${\rm Re}_{\rm m}=300$.} \label{tab1}
\end{center}
\end{table}

Table (\ref{tab1}) shows the estimated values of the
average work and dissipation over the $89\%$ of the computational
volume.
On the one hand, the dissipation is very weak (less than $0.12\%$
of the maximum dissipation level) and on the other hand, almost
all the net work occurs in those $89\%$ of the
volume. Moreover, the largest amount of dissipation and work occur
in the remaining $11\%$ of the volume and they are in almost
perfect detailed balance.

The above results lead to the conclusion that dynamo action occurs
primarily in regions where the field is weak. The net work is much
higher than the dissipation and as a result the field is amplified
by dynamo action. Most of the net work occurs when the weak field
is bent and stretched.
The magnetic energy is then distributed over the volume through the Poynting
flux and the field lines are dragged out and pile up against the local flux
cigars where balance occurs between stretching and diffusion.
The weak field is almost perfectly advected with the fluid and this
advection of the weak field would be possible even in the saturated regime
where the flow still has a good grip on the weakest field, although the
strongest field is now exerting a strong resistance against
{motion/stretching}, through the Lorentz force.

The above picture of the dynamo process {is most likely related to} the
observation that the growth rates found for the kinematic ABC dynamos are
always consistent with the large scale turn over times, even though the
ratio of the magnetic energy
to the rate of Joule dissipation (${\rm E}_{\rm mag} / {\rm Q}_{\rm J}$),
which is the time over
which the magnetic energy is destroyed and replenished, is much shorter.

In conclusion, the work done on the weak part of the field is
responsible for the increase in the magnetic energy by bending and
stretching the magnetic field lines. The exponential amplification
dominates the Joule dissipation which is weak through
almost all of the volume. The detailed balance between the Lorentz
work and the dissipation occurs only around the points where
strong magnetic field structures are formed. Thus, a dynamo
appears to be a process where amplification of the weak field
plays the dominant r\^{o}le in the growth of the magnetic energy.

\section{Conclusions}
\label{conclusions.sec}

In this paper turbulent dynamo action has been studied and some
new aspects of kinematic dynamo action have been illustrated.  It
was found to be important to keep the kinetic energy level of
the same order through the simulation, by automatically adjusting
the amplitude of the driving.  Using relatively
high values of ${\rm Re}_{\rm m}$ and ${\rm Re}$ we find that the
growth rate of the magnetic energy amplification {apparently} depends on the
stretching ability of the flow rather than on the values of the
above parameters. This is obvious even in the turbulent phase, for
different values of the magnetic Prandtl number, where the
dissipation is increased. The level at which the magnetic energy
saturates increases slightly with ${\rm Re}_{\rm m}$ but the
transient in the saturated regime is expected to become
insignificant at sufficiently high ${\rm Re}_{\rm m}$.

An important result is the existence of a near balance between the
Lorentz work and the Joule dissipation, which is apparent in both the
laminar and the turbulent regime of the forced ABC flow. Remarkably, the
balance originates primarily from small regions where strong magnetic flux
structures are concentrated. The nature of the dynamo is such that the
net growth of magnetic energy comes about through stretching and folding
of the weak magnetic field in the rest of the volume, where dissipation is
very weak.

The aim of this paper was not to study dynamo action on a global scale
(including aspects such as differential rotation, stratification and
convection) but rather to study the different stages of dynamos produced
by a family of hydrodynamically unstable flows.
However, the physical processes and the dynamo mechanisms studied
here are likely to be of similar importance in more realistic
stellar dynamo settings.

\begin{acknowledgements}
VA thanks the EU-TMR for support through a Marie Curie Fellowship.
SBFD was supported by the Danish Natural Science Research Council.
The work of {\AA}N was supported in part by the Danish National
Research Foundation, through its establishment of the Theoretical
Astrophysics Center.  Access to computational resources granted by
the Danish Center for Scientific Computing is gratefully acknowledged.
\end{acknowledgements}

\end{document}